# Microstructure manipulation by laser-surface remelting of a full-Heusler compound to enhance thermoelectric properties

*Leonie Gomell[a]\*, Tobias Haeger[b,c], Moritz Roscher[a], Hanna Bishara[a], Ralf Heiderhoff[b,c], Thomas Riedl[b,c], Christina Scheu[a], Baptiste Gault[a,d]\**

[a] *Max-Planck-Institut für Eisenforschung GmbH, Max-Planck-Str. 1, 40237 Düsseldorf, Germany*

[b] *Institute of Electronic Devices, University of Wuppertal, Rainer-Gruenter-Str. 21, 42119 Wuppertal, Germany*

[c] *Wuppertal Center for Smart Materials & Systems, University of Wuppertal, Rainer-Gruenter-Str. 21, 42119 Wuppertal, Germany*

[d] *Department of Materials, Royal School of Mines, Imperial College London, London, UK*

*\* corresponding authors. E-mail addresses: l.gomell@mpie.de (L. Gomell), b.gault@mpie.de (B.Gault)*

*Abstract*

There is an increasing reckoning that the thermoelectric performance of a material is dependent on its microstructure. However, the microstructure-properties relationship often remains elusive, in part due to the complexity of the hierarchy and scales of features that influence transport properties. Here, we focus on the promising Heusler-$Fe_2VAl$ compound. We directly correlate microstructure and local properties, using advanced scanning electron microscopy methods including in-situ four-point-probe technique for electron transport measurements. The local thermal conductivity is investigated by scanning thermal microscopy. Finally, atom probe tomography provides near-atomic scale compositional analysis. To locally manipulate the microstructure, we use laser surface remelting. The rapid quenching creates a complex microstructure with a high density of dislocations and small, elongated grains. We hence showcase that laser surface remelting can be employed to manipulate the microstructure to reduce the thermal conductivity and electrical resistivity, leading to a demonstrated enhancement of the thermoelectric performance at room temperature.

Additive manufacturing (AM) techniques are known for their capacity to shape components [1–4], but also to enable tailoring of the microstructure and therefore properties of a material on a fine scale [5–7]. Property manipulation can be achieved through the introduction of crystalline imperfections, i.e., structural defects [8] and compositional inhomogeneity, including the formation of secondary phases for instance [8–10] or the segregation of certain alloying elements [11–13]. Understanding the microstructure-property relationships is a key challenge for effective defect engineering. A promising application of microstructure engineering is the optimization of thermoelectric (TE) performance, which is typically revealed through the TE figure of merit $zT$ ($zT = \frac{S^2}{\rho\kappa}T$, with $S$: Seebeck coefficient, $\rho$: electrical resistivity, $\kappa$: thermal conductivity, and $T$: temperature). This $zT$ depends not only, but substantially on the inverse product of the electrical resistivity and the thermal conductivity [14]: ideal TE materials scatter phonons while conducting electrons [15]. Therefore, understanding and controlling transport properties of electrons and phonons is crucial when optimizing the TE properties of a given material [16].

To effectively scatter phonons of different wavelengths, a hierarchical microstructure with defects on multiple length scales is needed [17]. Such defect structures might include grain boundaries, dislocations, antisite defects, etc. A typical microstructure of casted material, which is characterized by comparatively slow solidification, does not include such a hierarchical defect structure. A more favorable microstructure can be generated by different synthesis pathways, including ball milling and plasma-spark sintering, and melt spinning. Post-processing, i.e., high-pressure torsion, can be used to enhance the TE properties by manipulating the microstructure [18–22]. AM also offers in principle the possibility to control the distribution and density of phonon scattering centers, as we recently demonstrated [23] by using laser surface remelting (LSR).

LSR has many similarities to additive manufacturing techniques. A focused laser selectively remelts the surface of the material, which is rapidly solidified afterwards by self-quenching [24]. The solidification rate and the maximum temperature can be controlled by adjusting the remelting parameters (i.e. laser power or laser scanning speed) [25,26]. Hence, it allows for manipulating the microstructure and control of the high defect density by adjusting the laser illumination parameters. Within the remelted region (called melt pool), quenching rates in the order of $10^3$ - $10^8$ K/s can be reached [27]. However, in contrast to AM techniques, no material is added through a powder feed, and only a bulk sample is needed.

Here, we studied the microstructure and transport properties of casted $Fe_2VAl$ manipulated by LSR. $Fe_2VAl$ exhibits a competitive combination of the Seebeck coefficient and the electrical resistivity (also called power factor ($S^2/\rho$) [14,28], which can compete with state-of-the-art materials such as $Bi_2Te_3$ [29]. However, its high thermal conductivity currently limits TE performance. Hence, the impact of additional scattering centers on the electrical resistivity and the thermal conductivity needs to be understood. We investigated the microstructure and its local influence on thermal conductivity and electrical resistivity. By combining multiple microscopy and microanalysis techniques on the same sample, we performed correlative, multiscale investigations of the microstructure, and in-situ measurements of properties. Our approach included scanning electron microscopy (SEM) and atom probe tomography (APT), as well as SEM-assisted four-point-probe technique and scanning thermal microscopy (SThM) to measure the electrical resistivity and thermal conductivity, respectively. We observed a complex arrangement of fine grains, containing a high dislocation density. This microstructure effectively scatters phonons, leading to a reduction of the thermal conductivity. At the same time, the electrical resistivity is reduced within the melt pool due to the off-stoichiometric composition. Our results demonstrate that we can improve thermoelectric performance by manipulating the microstructure using laser surface remelting, opening the path for possible thermoelectric device fabrication through AM, with local manipulation of the properties, with materials that are non-toxic and earth-abundant.

## Results and Discussion
### Microstructural analysis at the microscale

The microstructure of the casted sample is characterized by slow solidification during furnace cooling, with a solidification speed in the order of 10 K/s. This leads to large spherical grains (not shown here). In contrast, LSR ensures a high solidification rate, changing the microstructure significantly. The average chemical composition of the casted material is $Fe_{50.09}V_{24.95}Al_{24.96}$ (at.%), determined by inductively coupled plasma optical emission spectrometry (ICP-OES). 0.03 at.% of carbon and 0.01 at.% of nitrogen impurities were found by infrared absorption measurement and melting under helium with a subsequent thermal conductivity measurement, respectively.

The remelting parameters, i.e. the laser scanning speed and laser power, determine the solidification speed and the maximum temperature within the melt pool [30,31]. The sample coordinate system is chosen in a way that the laser scans in the positive z-direction, with $\vec{z}$ being the normal vector of the cross-sectional view. $\vec{y}$ is the normal vector of the remelted surface, with $y = 0$ μm marking the surface, and $\vec{x}$ is the normal vector of the side view (Fig. 1a). Fig. 1b-d shows a comparison of the melt pool cross-sections imaged by backscattered electrons (BSE), which were remelted using laser scanning speeds of 200 mm/s, 600 mm/s, and 1400 mm/s, respectively. Adjusting the scanning speed changes several melt pool characteristics, such as size, shape, and grain size [25,32,33]. Fig. 1e is a plot of the width of the melt pool as a function of the laser scanning speed. A decrease in the melt pool size with increasing scanning speeds is observed. The widths are determined from light-optical micrographs showing the top view of the sample, which are shown in the supplementary information (Fig. S1). When the scanning speed was increased from 50 mm/s to 2000 mm/s, the width decreased from $507 \pm 7$ μm to $97 \pm 1$ μm. The depth of the melt pool, analyzed by cross-sectional SEM images (not shown here), decreased from $285 \pm 10$ μm to $54 \pm 5$ μm for 50 mm/s and 2000 mm/s scanning speed. This size effect can be attributed to the shorter interaction time of the laser beam and the material, which leads to a decrease of the energy input [31,33]. Cracks are often observed at high-angle grain boundaries (HAGBs) of the casted material, if the HAGB is in the vicinity of the melt pool. Hence, the larger the melt pool is compared to the grain size of the casted sample, the higher is the chance of finding a HAGB in the heat-affected zone (HAZ), which is prone to cracking.

The depth-to-width ratio is a characteristic, which is often used to describe the melt pool mode, i.e. keyhole mode and conduction mode [26]. The cross-sectional BSE images (Fig. 1b-d) show a cross-over between keyhole mode (Fig. 1b) and conduction mode (Fig. 1d) for increasing scanning speed. The keyhole mode is characterized by a V-shaped melt pool, which is due to a high laser energy intensity and vaporization of the melt [26]. The resulting melt pool has a high depth-to-width ratio, i.e. 1.1 for a scanning speed of 200 mm/s. In turn, the conduction mode is characterized by a local melting process without significant vaporization, leading to a u-shaped melt pool and a smaller depth-to-width ratio, i.e. 0.6 for a scanning speed of 1400 mm/s.

We now focus on the sample with a laser scanning speed of 1400 mm/s, which showed a good combination of structural integrity and fine-scaled microstructure. For this sample, we combine information obtained from the different viewing directions, i.e. the cross-section view (*xy*-plane), the side view (*yz*-plane) approximately in the center of the melt pool, and the top view (*xz*-plane). The top of the sample was imaged without any polishing and at $y \approx 5$ µm and $y \approx 20$ µm after grinding and polishing. In the following, these samples will be referred to as $top_0$, $top_5$, and $top_{20}$, respectively. In all figures, the imaged regions with regards to the melt pool are indicated by a sketch.

Elongated grains are observed in the cross-section and the side view (Fig. 1d and Fig. 2). These grains grow from the HAZ in the direction of the maximal temperature gradient, which changes across the melt pool depth and is tilted with respect to the surface normal [34,35]. Hence, the grains in the side view are slightly tilted in the direction of the laser movement. In the central plane of the melt pool, the growth rate $V_S$ is related to the laser scanning speed $V_B$ by $V_S = V_B \cdot \cos \theta$, where $\theta$ is the angle between $V_S$ and $V_B$, which is a function of the position within the melt pool [36]. $\theta$ is evaluated in the bottom 20 µm of the melt pool and the top 10 µm. The grains near the surface are too small and not elongated enough for a reliable analysis of the angle.

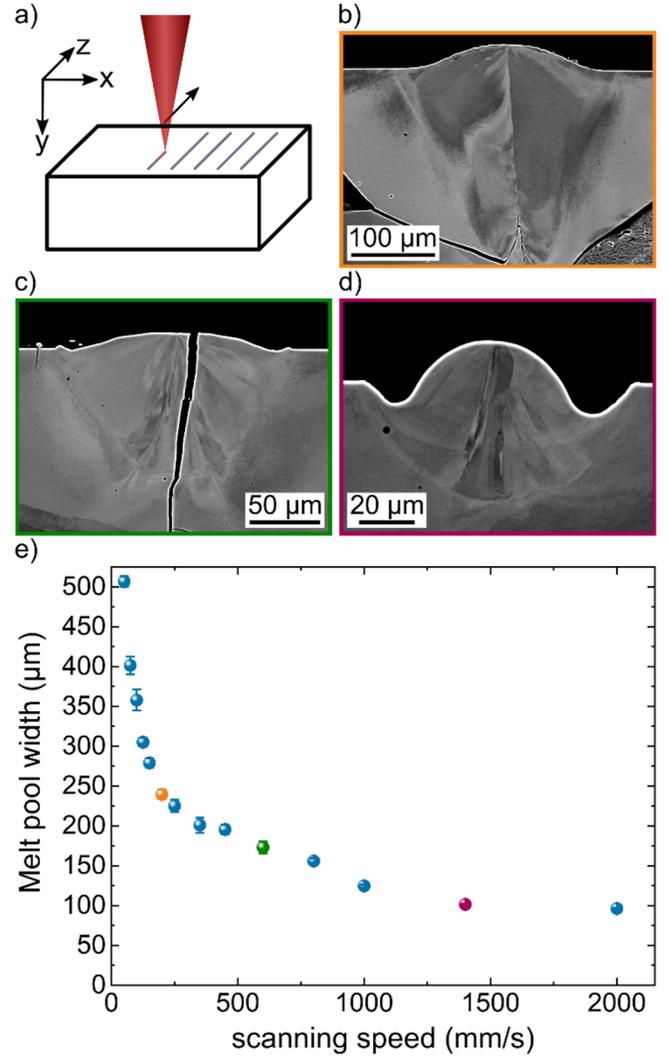

*Fig. 1a) Schematic of the LSR setup, indicating the chosen sample coordinate system. b-d) BSE images of melt pools with increasing laser scanning speeds of b) 200 mm/s, c) 600 mm/s, and d) 1400 mm/s. e) Melt pool width as a function of laser scanning speed. The width was determined by optical top view images, which are displayed in Fig. S1 (ESI). The orange, green and purple dots are showing the width for remelting conditions used in b, c, d, respectively.*

The grains close to the HAZ are inclined by $\theta = 80° \pm 3°$, leading to a solidification speed of $V_S = 1400 \text{ mm} \cdot \text{s}^{-1} \cdot \cos 80° \approx 243 \text{ mm} \cdot \text{s}^{-1}$. In the top 10 µm of the melt pool, the solidification speed increases, indicated by a decrease in $\theta$. Here, $\theta = 75° \pm 2°$ is observed, leading to a solidification speed of approx. $V_S = 1400 \text{ mm} \cdot \text{s}^{-1} \cdot \cos 75° \approx 362 \text{ mm} \cdot \text{s}^{-1}$.

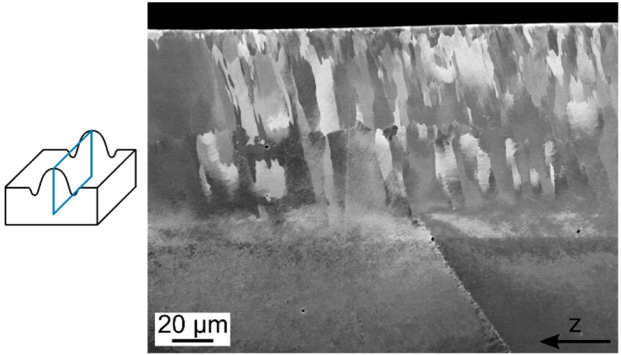

*Fig. 2. BSE image of the side view of the melt pool remelted with a laser scanning speed of 1400 mm/s. The schematic on the left side shows the location of the imaged region with regard to the melt pool. The z-arrow indicates the laser scanning direction.*

Closer to the surface, i.e. in the top 2 µm, it is not possible to obtain the growth direction, due to a lack in elongation of the grains. The height, width, and depth of the grains in the center of the melt pool are approximately 14 ± 5 µm, 2 ± 1 µm, and 3 ± 2 µm, respectively, as evaluated from all viewing directions. Some grains span the entire depth of the melt pool. At the bottom and on the sides of the melt pool, the grain size is larger due to the slower solidification rate [30,32,37,38].

Fig. 3 shows top view micrographs of the melt pool in the $top_0$, $top_5$, and $top_{20}$ samples. The different positions in the melt pool are analyzed to observe the impact of the varying solidification speed within the melt pool on the grain structure. BSE images reveal a columnar grain growth in the direction of the laser movement. The boundaries of the grains/columns are irregular and wavy. Fig. 3c ($top_{20}$ sample) shows a GB of the casted sample crossing the melt pool. The GB is nearly perpendicular to the melt pool and straight on the scale of the melt pool width. Within the melt pool, the GB forms a V-shape with an angle of approx. 155°, deviating from the straight line. This form is associated with the three-dimensional growth direction of the grains.

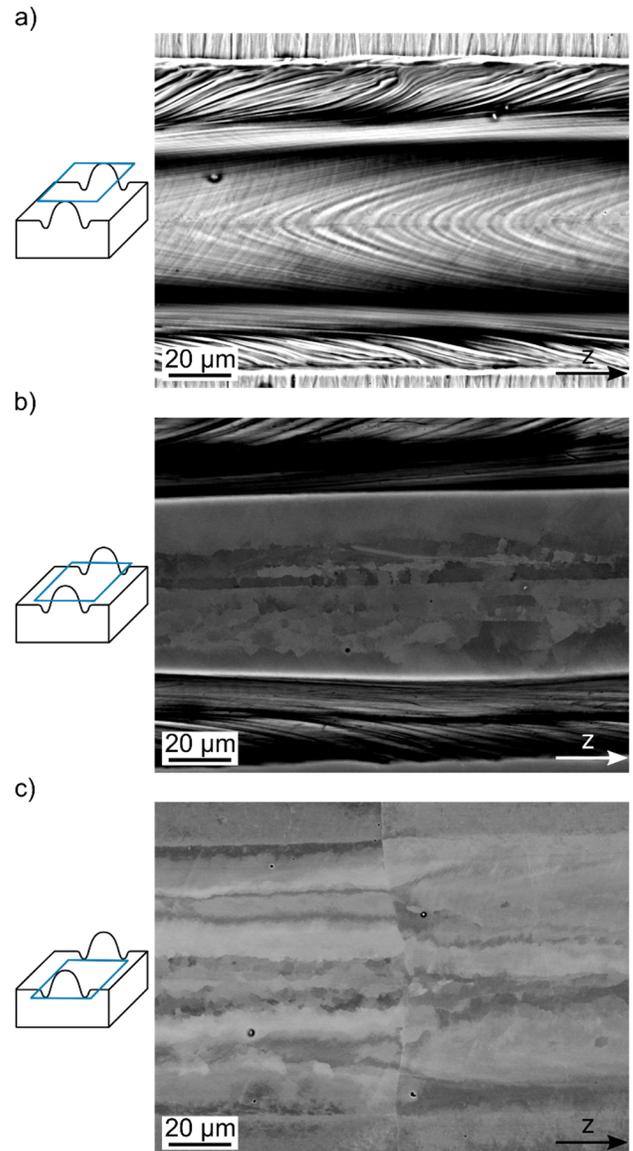

*Fig. 3. BSE images of the top view of the melt pool remelted with a laser scanning speed of 1400 mm/s. The approximate position within the melt pool is indicated in the sketches on the left of the BSE images. a) as produced, without polishing ($top_0$ sample), b) after grinding and polishing of approx. 5 µm ($top_5$ sample), and c) after grinding and polishing of approximately 20 µm ($top_{20}$ sample). A high-angle grain boundary of the casted material is visible in the center of the image crossing the melt pool perpendicularly. The z-arrow indicated the laser scanning direction.*

Hence, the grains tilt towards the scanning of the grains. Hence, the grains tilt towards the scanning direction and the HAGB appears in V-shape when cut in a central position of the melt pool.

The crystallographic orientation of the grains is analyzed using electron backscatter diffraction (EBSD). The casted sample shows an untextured grain structure with HAGBs (Suppl. Fig. S1d of Ref. [23]). Within the melt pool, no new HAGBs are formed, but newly formed grains are separated by low-angle grain boundaries (LAGB) due to the epitaxial growth on the substrate grain. EBSD maps of the cross-section, the side view, and top$_5$ and top$_{20}$ samples are shown in Fig. 4a-d, respectively. The grains within the melt pool grow epitaxially from the HAZ, leading to a strong preferred orientation, visible as a minor color/orientation contrast in the EBSD inverse pole figure. For all maps, the inverse pole figure is shown in the direction of $\vec{z}$, which is the laser scanning direction. Due to a similar orientation of the substrate grain, all inverse pole figure maps show a reddish color, indicating a [001] growth direction. Yet, this is attributed to the substrate grain and not to preferential growth, as melt pools within different substrate grains show a different crystallographic orientation (Fig. S2 (ESI)). This is contrary to several studies of laser remelted materials with a cubic crystal structure [25,34,39], where the [001] growth direction was favored. Kurzynowski et al. discussed that the growth direction depends on the scanning speed. While slow scanning favors directional solidification, fast scanning leads to epitaxial growth without any preferred growth direction [25]. Yet, we observe epitaxial growth also for slower remelted melt pools with a scanning speed of 200 mm/s and 600 mm/s (Fig. S2 b, c (ESI)). The misorientation can be visualized by Kernel average misorientation maps and misorientation maps. The former shows the misorientation between two adjacent points on a scale of 0°-2°, while the latter shows the misorientation relative to a point in the substrate grain on a scale of 0°-10°. The grains in the central path of the melt pool are most strongly misoriented with regards to the substrate grain and with regards to neighboring grains. The misorientation between the HAZ and the melt pool is in the order of 1°.

The chemical composition of the grains was determined using energy-dispersive X-ray spectroscopy (EDX) and electron probe microanalysis (EPMA). The casted sample shows a homogeneous elemental distribution (Supplementary Fig. S1 in Ref. [23]). On the contrary, after LSR, a depletion of Al, compensated by an enrichment of V is observed within the melt pool (Fig. 5). Further, the composition is not homogenous within the melt pool, but a region with a stronger increase of V and a decrease of Al is observed, which is attributed to banding. The band formation is observed in all EDX maps (Fig. 5), but best visible in the cross-sectional map (Fig. 5a), where it forms a half-ring shape from the edges of the melt pool towards the center. The BSE image of this region (Fig. 1d) shows a slightly brighter mass contrast, underlining that the V-rich region is within the melt pool and not at its edge. In the top view, the V-rich region is cut, leading to V-rich and Al-poor bands along the laser scanning direction.

The compositional change across the melt pool was quantified by EPMA (Fig. 5d-f). The scan direction is indicated by the white arrows. In the band, the composition reaches $Fe_{49.8\pm0.7}V_{27.7\pm0.7}Al_{22.5\pm0.7}$ (at.%) as determined in the cross-sectional measurement and confirmed by the other two measurements. This composition can be expressed as $Fe_2V_{1.1}Al_{0.9}$. Within the melt pool closer to the impact position of the laser (in regions with a higher maximal temperature), the slight decrease in Al and an increase in V ($Fe_{49.5\pm0.7}V_{26.0\pm0.7}Al_{24.5\pm0.7}$ (at.%)) is observed compared to the stoichiometric composition of the casted material. According to the phase diagram calculated by Berche et al. [45], this composition is outside the equilibrium solubility range of $L2_1$ $Fe_2VAl$. Yet, no secondary phase could be observed using APT and EBSD, and a non-equilibrium composition can be achieved because of the fast cooling rate.

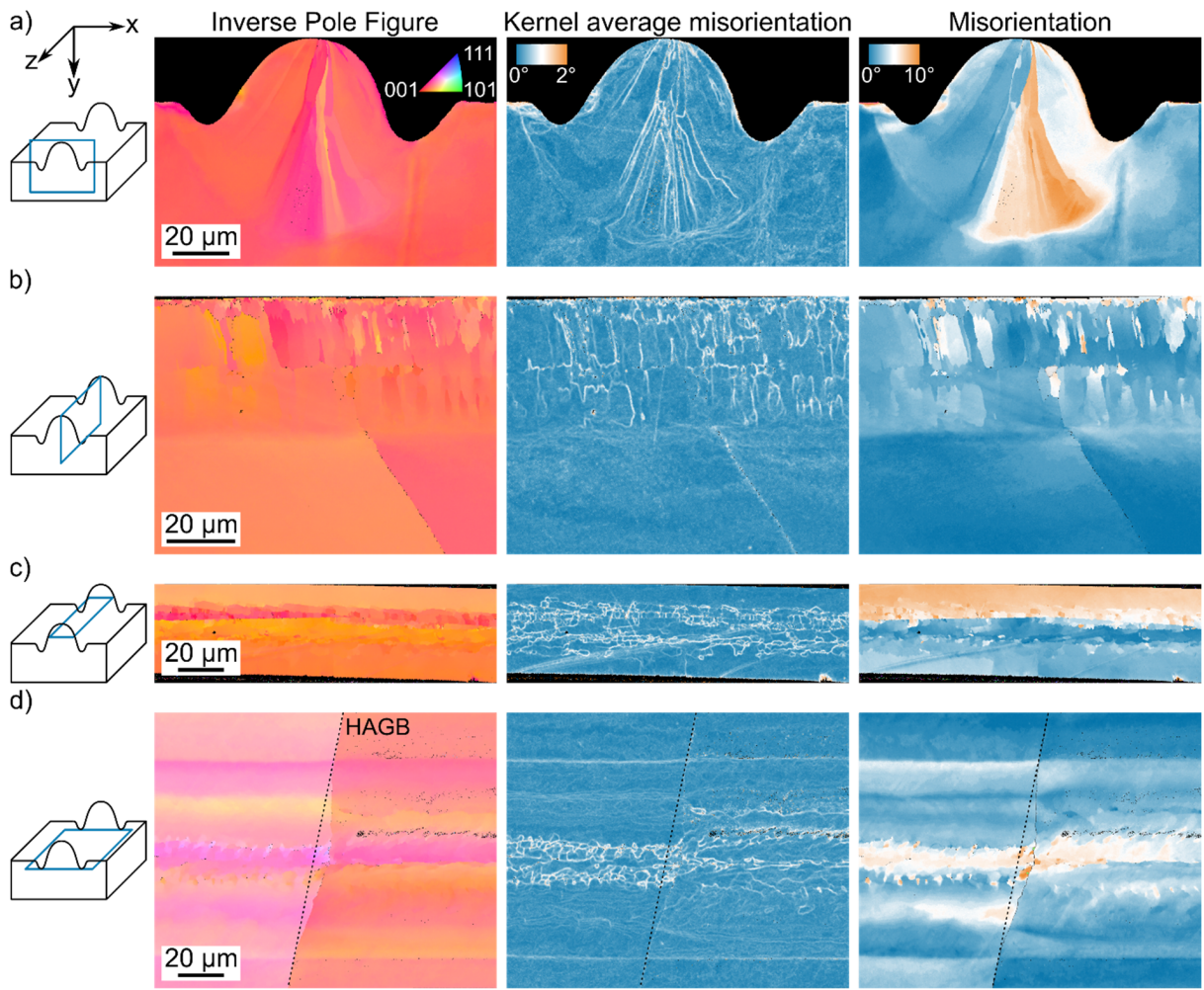

*Fig. 4) Investigation of the crystallographic orientation within the melt pool by EBSD measurements. The sketches indicate the position of the scan within the melt pool. Left column: inverse pole figure. For all viewing directions, the direction was chosen in [001] direction, hence in laser scanning direction. Epitaxial growth of grains within the melt pool is observed. Middle column: Kernel maps, visualizing misorientations between the second nearest neighbor on a scale of 0°-2°. Right column: Misorientation map with regard to the substrate grain on a scale of 0°-10°. a) cross-sectional view, b) side view, c) $top_5$ sample, d) $top_{20}$ sample.*

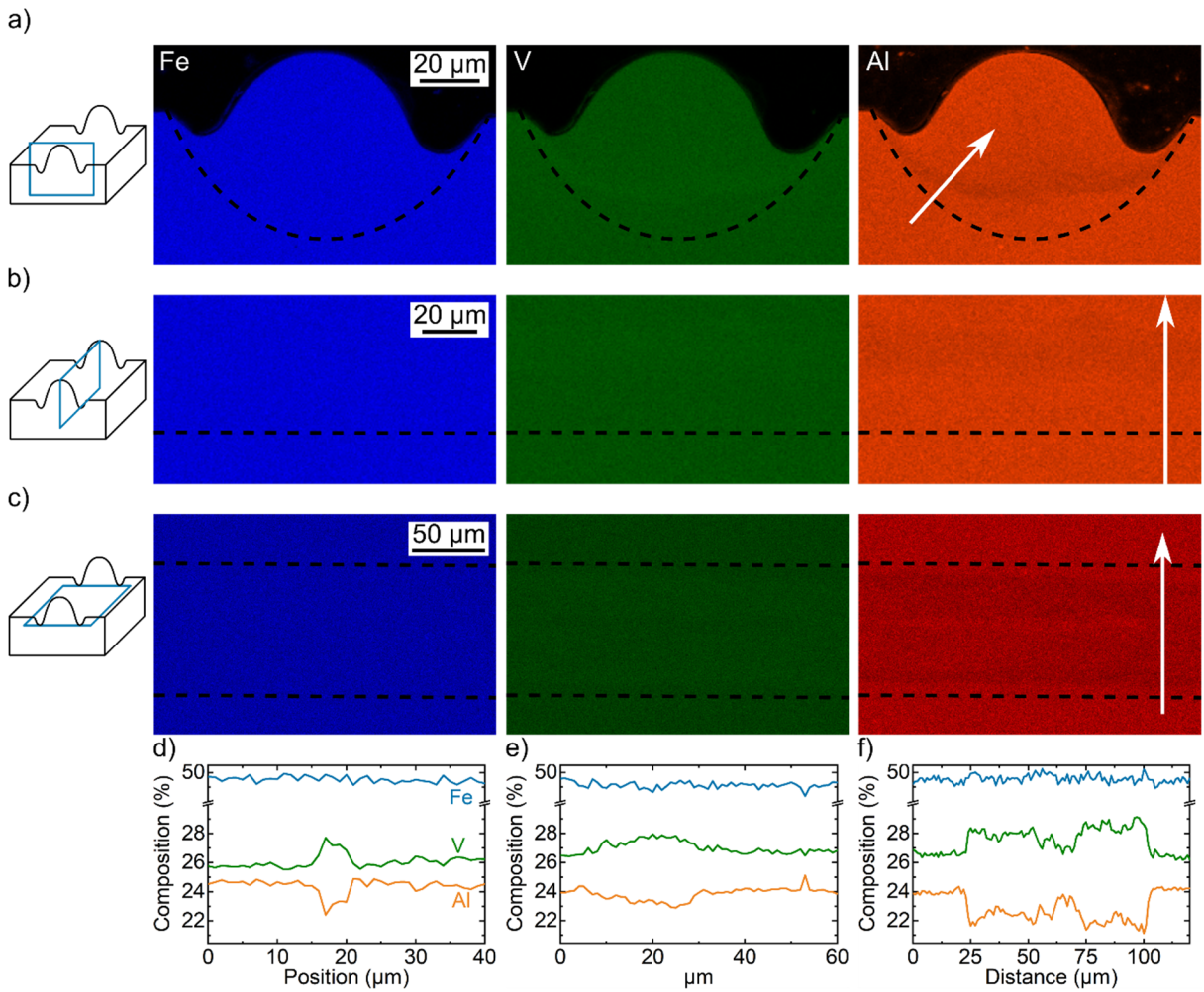

*Fig 5. Compositional analysis by EDX and EMPA of the melt pools in the different viewing directions. The melt pool boundaries are indicated by the dashed line. A V-rich and Al-poor region is observed within the melt pool. The sketches on the left side indicate the position within the melt pool. a) Cross-section, b) side view, and c) top view (top$_{20}$ sample). The HAGB of the casted sample (Fig. 2c) is central in the maps. No indication of any segregation in the micrometer scale towards the HAGB is observed. d, e, f) EPMA line scans across along the direction of the white arrow in a, b, c.*

The simplest explanation for the loss of Al is preferential evaporation, as Al has the lowest melting point of the constituents of the material. One expects an increase in evaporation at higher temperatures. Yet, we only observe an off-stoichiometric composition and band formation in melt pools, remelted with a laser scanning speed above 1000 mm/s. For scanning speeds between 50 mm/s and 800 mm/s, a homogenous elemental distribution is observed, even if the maximum temperature is increased. We suggest two possible reasons for such a behavior: First, the faster remelting leads to an intensified Marangoni flow within the melt pool [35], which can remove any protective oxide layer and brings new Al to the top of the melt pool. Hence, more Al can evaporate in the faster remelted regions compared to the slower remelted regions with fewer turbulences in the liquid state, where a protective Al-oxide layer can be build up. Bormann et al. used the same argument to explain their observation of increased Ni evaporation in a NiTi alloy [40]. Another explanation would be a stronger spattering for higher scanning speed [41], with a preferential spatter of Al due to the lower melting temperature of Al compared to Fe and V.

The additional V enrichment is explained by solute trapping due to the fast solidification speed. Band formation occurs, which leads to an even stronger V-enriched/Al-depleted zone. Such banding is known from welding and laser surface remelting literature [25,42]. It occurs due to a local change of the solidification rate $R$, which fluctuates cyclically above and below the mean growth rate [43], combined with the Marangoni flow within the melt pool [32]. A rapid increase of $R$ can change the solubility of the solvent in the solution, causing excess dissolved elements to be deposited in the solid. This results in a local change of the composition and can promote antisite V/Al defects [44].

Not only boundaries but also other defects such as dislocations can affect the electrical and thermal conductivity due to their scattering ability. ECCI is used to image these defects in bulk materials [46]. ECC images of the casted sample and the LSR sample reveal a significant change in the defect density (Fig. 6). The casted sample shows hardly any dislocations (Fig. 6a). Only two-dimensional features are found, likely to be stacking faults, oriented parallel and perpendicular to each other, with a number density of approx. $8 \cdot 10^{10}$ m$^{-2}$.

The LSR manipulated material in top view (top$_0$ sample) is imaged in Fig. 6b. The dislocations appear to be oriented in a similar direction until they reach the surface. The defect density is approximately $6.8 \cdot 10^{13}$ m$^{-2}$, i.e. three orders of magnitudes higher compared to the casted material. A LAGB is observed, composed of several dislocations (white arrows). The defect structure in the cross-section after OPS polishing is shown in Fig. 6c. The image was taken approximately in the center of the melt pool, and thus, at a position of a lower solidification speed compared to Fig. 6b. The position is indicated in Fig. S3a (ESI). The observed dislocation density is approximately $3.3 \cdot 10^{13}$ m$^{-2}$ and therefore slightly lower, but in the same order of magnitude compared to the impact position of the laser. A close-up, which was used to calculate the dislocation density, is presented in Fig. S3b (ESI). Partly, dislocation pairs, i.e. dislocation dipoles are observed. An additional ECC image is shown in the supplementary Fig. S3c, showing a low-angle tilt grain boundary with parallel dislocations. Also, low angle twist boundaries have been observed with dislocation pairs forming a V-shape. Scanning transmission electron microscopy (STEM) confirms the high dislocation density, as shown in Fig. S4, albeit with a relatively smaller probed volume and lower statistics compared to ECCI.

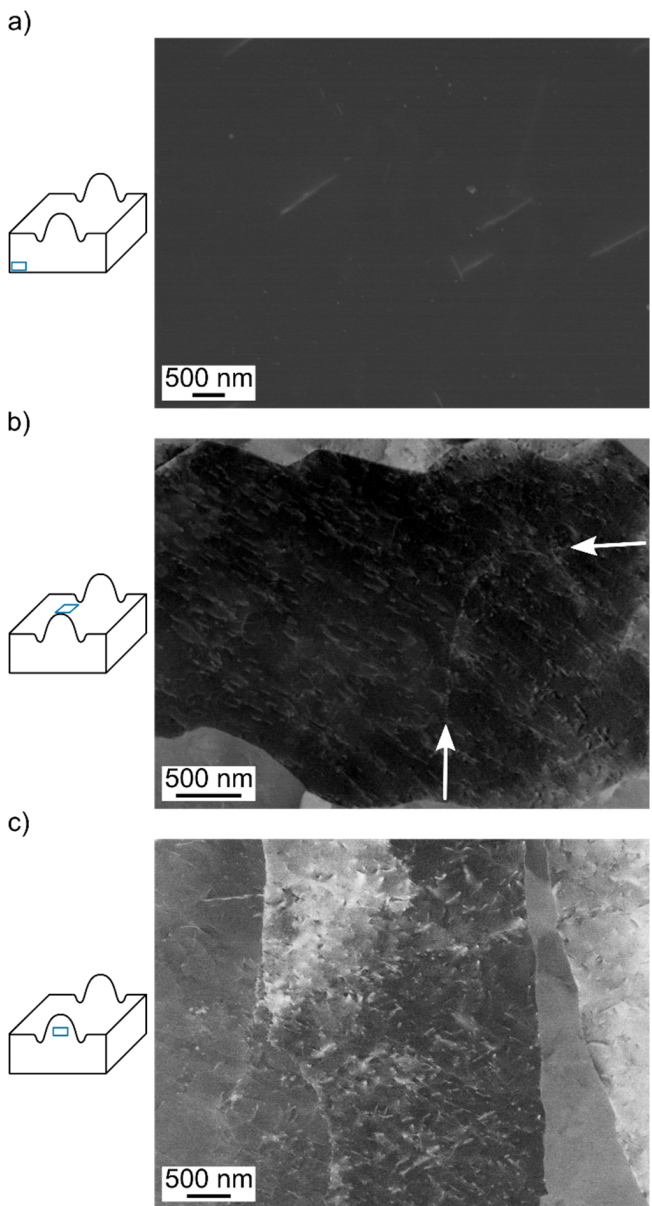

*Fig 6. ECC images, showing defects on the nanometer scale. a) Casted sample, b, c) LSR sample, with b) unpolished top view (top$_0$ samples), and c) cross-section. The white arrows in b) indicate the position of a LAGB. The location within the melt pool and an image with higher magnification for the determination of the dislocation density is given in the supplementary information (Fig. S3).*

**Atom probe tomography**

APT was performed in a region close to the impact position of the laser and in the V-rich zone of the melt pool. Close to the impact position, seven APT specimens were run successfully, leading to more than 500 million detected ions. The average composition is $Fe_{50.0\pm0.3}V_{25.6\pm0.3}Al_{24.3\pm0.2}$ (at.%). The remaining 0.1 at.% are impurities, such as C, N, Ar, B, and Si. Compared to the composition of the casted material, which was obtained by ICP-OES, a slight enhancement in V and a slight depletion in Al is observed, in accordance with the EPMA and EDX compositional data. Species loss of Al could also arise from the APT measurement itself due to its lower average evaporation field [47–49], but this was not observed in the APT analysis of the as-cast material (not shown).

Elemental segregation to dislocations and grain boundaries is observed. GBs were found in two APT data sets (Fig. 7a-c and Suppl. Fig. S5), indicated by elemental segregation and a shift in the position of the main crystallographic pole [50]. The detector hit maps are used to estimate the misorientation between the two grains by comparing the positions of the pole before and after the boundary (Fig. 7c and Fig. S5c), translating into a misorientation of approximately 5.7° and 1.2°, respectively, assuming a field of view of 60° on this instrument [51]. Thus, the observed boundaries can be classified as LAGB, equivalently to the EBSD and ECCI results.

Nitrogen segregation was found at both boundaries, as indicated by the iso-composition surfaces in Fig. 7a and Fig. S5a. The iso-composition surface marks regions where the nitrogen concentration is higher than 0.3 at.%. The observed nitrogen was detected as part of $VN^+$ or $VN^{2+}$ molecular ions typically detected in (carbo)nitrides precipitates in Fe-based materials [52,53]. The elemental composition across the LAGB is analyzed using a one-dimensional composition profile (Fig. 7b and Fig. S5b) within a 30-nm-diameter cylindrical ROI. The step size for the calculation was set to 0.3 nm. At the boundary, N is enriched up to 0.2 at.%, which is an order of magnitude higher than the nitrogen composition in the matrix (0.02 ± 0.01 at.%, determined within the same dataset). In addition, a minor enrichment of V is found, together with a minor depletion of Al. No change in the Fe composition is observed. The N segregation forms a loose pattern made of several lines within the boundary, pointing to low-level segregation to dislocations forming a LAGB that is difficult to evidence.

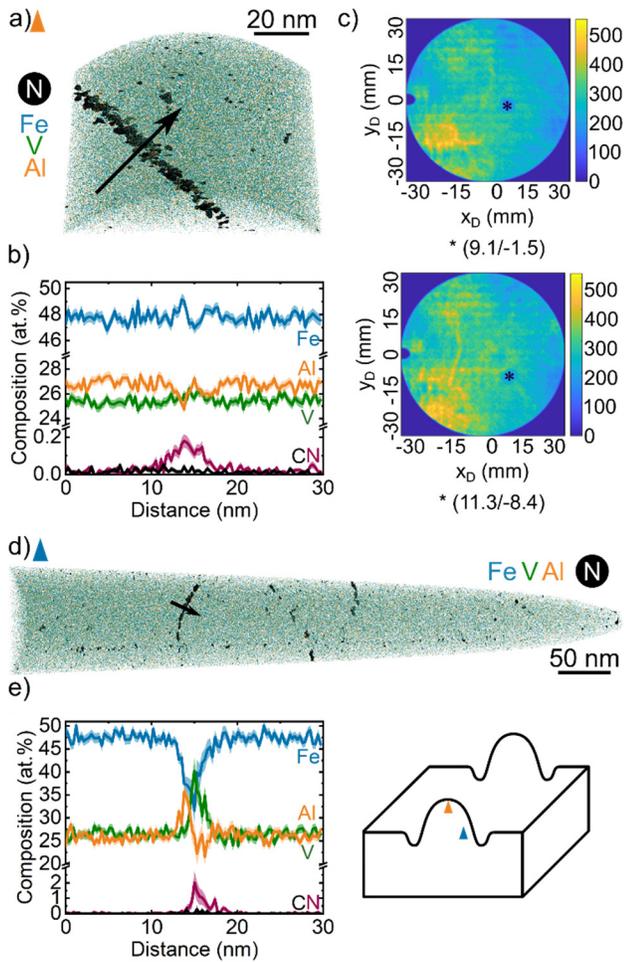

*Fig. 7. APT reconstruction of $Fe_2VAl$ after LSR. The specimen shown in a-c is taken close to the impact position of the laser (indicated by the orange triangle in the sketch in the bottom right corner), the second specimen shown in d-e is taken from the V-rich region (indicated by the blue triangle). a, d) APT reconstruction, showing the matrix atoms Fe, V, and Al in blue, green, and orange. The black iso-composition surface shows regions with a nitrogen concentration of more than 0.3 at.%. b, e) compositional profile across the LAGB and dislocation. An enrichment of N and V is observed, while Al is slightly reduced. c) Detector hit maps on either side of the GB showing the positions of the (001) pole approximately in the center of the map. The position of the pole is used to estimate the misorientation at the GB.*

Segregation is also found at dislocations inside the grains (Fig. 7d-e), here shown within an APT specimen prepared from the V-rich zone of the melt pool. The enrichment of V and C and the depletion of Fe and Al is stronger than in the region close to the impact position of the laser. A 1D composition profile is obtained across the dislocation in a cuboidal region of interest (ROI) (20 x 4 x 30) nm³. At the dislocation, the concentration of V increases up to 40 at.%. Nitrogen is enriched up to more than 2 at.%, while Fe is depleted to approximately 34 at.% and Al to 24 at.%. Next to the dislocation, an overshoot of Al is observed, indicating dislocation movement [54,55]. The increased segregation to this dislocation compared to the dislocations within the LAGB can be attributed to a stronger solute trapping in the banding affected region.

Segregation of nitrogen has been observed at boundaries in $Fe_2VAl$ before after laser surface remelting and melt spinning [23,56]. In both reports, the segregation was more pronounced compared to the present report. Ref. [23] analyzed laser surface remelting of the surface area with a hatch distance of 0.1 mm. This leads to an overlap of neighboring scan tracks and intrinsic heat treatment, which can increase the segregation by activating the diffusion of the impurity atoms during remelting of the subsequent tracks. The intrinsic heat treatment is avoided in the current report, which reduces the level of segregation. Also, the faster laser scanning speed used here (1400 mm/s vs. 350 mm/s in Ref. [23]) can reduce the level of segregation.

**Thermal conductivity and electrical resistivity**

Fig. 8 shows local measurements at room temperature of thermal conductivity and electrical resistivity measured at different positions inside and outside the melt pool. As the composition plays a crucial role in carrier transport, the local vanadium composition determined by EPMA is also plotted in the figure. As shown in Fig. 5, the excess V is compensated by Al depletion while the Fe composition is approximately constant.

Outside the melt pool, the thermal conductivity is 27.3 ± 0.4 W/mK, and the electrical resistivity is 6.9 ± 0.2 µΩm. Both values agree with previous reports of casted $Fe_2VAl$ [57–60]. Inside the melt pool, the thermal conductivity monotonously decreases down to 26.2 ± 0.3 W/mK. The electrical resistivity increases to 8.0 ± 0.2 µΩm at the edge of the melt pool and then decreases monotonously to 6.8 ± 0.2 µΩm measured in the center of the melt pool. The decrease in thermal conductivity is attributed to the high density of defects: phonons are scattered by dislocations and antisite defects associated with the off-stoichiometric composition. The electrical resistivity is also increased by these scattering centers, which are observed at the edge of the melt pool. However, within the melt pool, the off-stoichiometric composition changes the carrier concentration and in turn reduces the electrical resistivity. Hence, towards the center of the melt pool, the electrical resistivity decreases. Such a decrease of $\rho$ was previously reported for off-stoichiometric $Fe_2V_{1-x}Al_{1+x}$ ($x<0$) compounds [61,62]. Furthermore, it is evident from the composition-electrical resistivity relations that the high concentration of dislocations has a negligible effect on the resistivity. For efficient scattering, the distance between the scattering centers needs to be shorter than the mean free path. For phonons, the mean free path is approx. 2.4 µm [56], while the mean free electron path is in the order of 10 nm. Hence, the observed population of structural defects exhibits stronger scattering for phonons compared to electrons. The increase of the TE performance can be estimated by comparing the product of the electrical resistivity and the thermal conductivity outside and inside the melt pool. An increase of approx. 6 % is observed here.

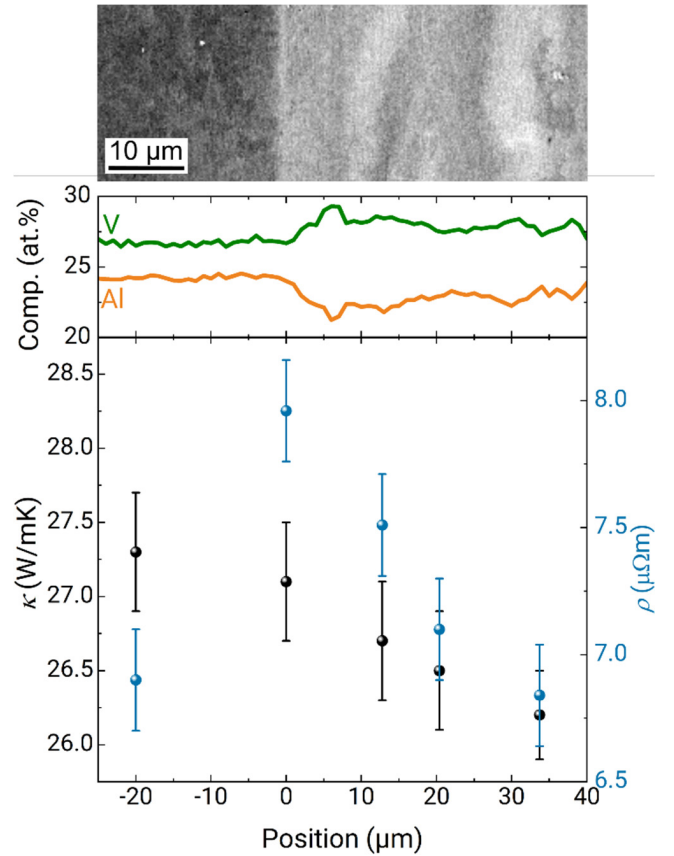

*Fig. 8: Measurement of the thermal conductivity $\kappa$ (black) and electrical resistivity $\rho$ (blue) at different positions within the melt pool. 0 µm marks the edge of the melt pool. The SEM image and the compositional changes across the melt pool for V and Al are given for scaling and comparison.*

To conclude, we measured the microstructure and correlated it with local measurements of the thermal conductivity and electrical resistivity. With this method, we can investigate the microstructure-property relationship of thermoelectric $Fe_2VAl$. SEM investigations showed epitaxial growth of elongated grains within the melt pool. Banding leads to an increase of the V composition, while Al is depleted. A high dislocation density is observed using ECCI and APT. Segregation of V and N to low angle grain boundaries and dislocations is observed, in agreement with earlier publications.

This microstructure leads to a decreased thermal conductivity and decreased electrical resistivity, attributed to the high density of scattering centers and the off-stoichiometric composition, respectively. Hence, LSR can be used to increase the thermoelectric performance of $Fe_2VAl$ by 6 % as shown here. Due to the similarity of the microstructure of LSR and selective laser melting, this study motivates the synthesis of selective laser melting built of legs for complete devices, for instance, with controlled defect densities in order to optimize the thermoelectric performance.

*Experimental Methods*

**2.1: Synthesis**

Stoichiometric amounts of pure Fe (99.9%, Carboleg GmbH), Al (99.7%, Aluminium Norf), and V (99.9%, HMW Hauner GmbH) were mixed and casted to synthesize $Fe_2VAl$. The sample was flipped over and remelted four times to ensure homogeneity

Subsequently to casting and after hand grinding to 600 grit SiC paper, laser surface remelting (LSR) was used to manipulate the microstructure. An AconityMini system (Aconity 3D) is used to remelt single lines with a distance of at least 0.5 mm. A pause of at least 90 seconds was allowed between remelting processes to let the sample cool down to room temperature, ensuring identical conditions for every melt pool. The setup includes an ytterbium-fiber laser with a wavelength of 1070 nm focused onto a 90 μm-diameter spot. The laser power was set to 200 W and the scanning speed was varied between 50 mm/s and 2000 mm/s. The remelting process was conducted in argon atmosphere with a residual oxygen concentration below 80 ppm.

**2.2. Experimental Methods**

We investigate the microstructure on the micrometer-to-nanometer scale using scanning electron micro-scopy (SEM) and related techniques such as electron backscattered diffraction (EBSD), energy-dispersive X-ray spectroscopy (EDX), and electron channeling contrast imaging (ECCI). Prior to imaging, the samples were mechanically polished down to 0.05 μm colloidal silica.

Backscattered electron (BSE) imaging and ECCI were conducted in a Zeiss Merlin SEM. For EBSD and EDX a Zeiss Sigma SEM was used. The acceleration voltage was set to 15 kV for EDX, 20 kV for EBSD, and 30 kV for ECCI. EBSD mapping was conducted with a step size of 200 nm on a hexagonal grid. Local compositional changes on the micron-scale were observed by electron probe microanalysis (EPMA) using a JEOL JXA-8100. A JEOL 2100Plus TEM was used for transmission electron microscopy.

Atom probe tomography (APT) was conducted for microstructural observations with sub-nanometer resolution. Needle-shaped specimens were prepared from positions close to the impact position of the laser and within the melt pool using a dual-beam focused-ion-beam (FIB) instrument (FEI Helios Nanolab600/600i) with a Ga ion source. The *in-situ* lift-out method and subsequent sharpening process are described in Ref. [63]. The APT specimens were analyzed using a LEAP™ 5000 XS instrument (Cameca Instruments Inc., Madison, WI, USA) operated in laser pulsing mode using a pulse repetition rate of 200 kHz and a pulse energy of 60 pJ. The base temperature of the specimen was kept at 60 K and the target detection rate was set to 1% or 2%. AP Suite 6.1 software (Cameca Instruments) was used for data reconstruction and analysis.

Local properties measurements of the electrical resistivity and the thermal conductivity were carried out on the top$_{20}$ sample on four different positions from outside to inside the melt pool. Electrical resistivity measurements were conducted using an *in-situ* 4-point-probe technique inside a Zeiss Auriga SEM. The probes have a tip radius of approx. 50 nm and can be positioned by four independent micromanipulators (PS4, Kleindiek Nanotechnik GmbH). The probes are positioned approximately 1 μm apart. Technical details on the electrical measurements, e.g., errors, corrections due to deviation from equidistance positions of needles, modulation of electrical current to decrease noise, are described in Ref. [64].

Measurements of local thermal properties were performed by scanning near-field thermal microscopy (SThM), using the so-called 3ω-technique [65, 66]. The technique is based on atomic force microscopy using a resistive probe (Bruker Vita-SThM) that is electrically powered with a frequency between 3 kHz and 7 kHz. More details of the SThM system can be found in Ref. [67]. The SThM is integrated into the analysis chamber of an environmental scanning electron microscope (ESEM), which allows selection of the region of interest by electron microscopy and concomitant analysis of the thermal properties by SThM under vacuum conditions.

Acknowledgements

We thank U. Tezins, A. Sturm, M. Nellessen, C. Broß, and K. Angenendt for their technical support at the FIB/APT/SEM facilities at MPIE. S. Zaefferer, E. A. Jägle, and G. Dehm are acknowledged for helpful discussions. L. G. gratefully acknowledges Studienstiftung des deutschen Volkes for funding. T. H. R. H., T. R. acknowledge funding by the Deutsche Forschungsgemeinschaft (DFG) under the project numbers HE2698/7-2. H. B. acknowledges the financial support by the ERC Advanced Grant GB CORRELATE (Grant Agreement 787446 GB-CORRELATE) of Gerhard Dehm.